\documentclass{aa}
\usepackage{natbib,psfig,graphicx,ulem}
\usepackage{txfonts}
\usepackage{longtable}
\usepackage{soul,color}

\def\mathnew{\mathsurround=0pt}
\def\simov#1#2{\lower .5pt\vbox{\baselineskip0pt \lineskip-.5pt
\ialign{$\mathnew#1\hfil##\hfil$\crcr#2\crcr\sim\crcr}}}

\def\MeV{Me\kern-0.11em V}
\def\keV{ke\kern-0.11em V}

\begin{document}

\title{Diffuse light in the young  cluster of galaxies CL~J1449+0856 at z=2.07
\thanks{The data presented in this paper were obtained from the Mikulski 
Archive for Space Telescopes (MAST). STScI is operated by the Association 
of Universities for Research in Astronomy, Inc., under NASA contract 
NAS5-26555. Support for MAST for non-HST data is provided by the NASA Office 
of Space Science via grant NNX09AF08G and by other grants and contracts. This 
research has made use of data from HST-COSMOS project, held in the HST-COSMOS 
database operated by Cesam, Laboratory of Astrophysics of Marseille.}}

\author{C.~Adami\inst{1} \and 
F.~Durret\inst{2} \and 
L.~Guennou\inst{1} \and
C. Da Rocha\inst{3}
}

\offprints{C. Adami \email{christophe.adami@oamp.fr}}

\institute{
LAM, OAMP, Universit\'e Aix-Marseille $\&$ CNRS, P\^ole de l'Etoile, Site de
Ch\^ateau Gombert, 38 rue Fr\'ed\'eric Joliot-Curie,
13388 Marseille~13 Cedex, France
\and 
UPMC-CNRS, UMR7095, Institut d'Astrophysique de Paris, 98bis Bd Arago, F-75014, Paris, France 
\and
N\'ucleo de Astrof\'{\i}sica Te\'orica, Universidade Cruzeiro do Sul, 
R. Galv\~ao Bueno 868, 01506--000 S\~ao Paulo, SP, Brazil
}

\date{Accepted . Received ; Draft printed: \today}

\authorrunning{Adami et al.}

\titlerunning{Diffuse light in the young cluster of galaxies
  CL~J1449+0856 at z=2.07}

\abstract 
{Cluster properties do not seem to be changing significantly during
  their mature evolution phase, for example they do not seem to show
  strong dynamical evolution at least up to z$\sim$0.5, their galaxy
  red sequence is already in place at least up to z$\sim$1.2, and
  their diffuse light content remains stable up to z$\sim$0.8. The
  question is now to know if cluster properties can evolve more
  significantly at redshifts notably higher than 1.  }
{We propose here to see how the properties of the intracluster light
  (ICL) evolve with redshift by detecting and analysing the ICL in the
  X-ray cluster CL~J1449+0856 at z=2.07 (discovered by Gobat et
  al. 2011), based on deep HST NICMOS H band exposures.}
{We used the same wavelet-based method as that applied to 10 clusters
  between z=0.4 and 0.8 by Guennou et al. (2012).}
{We detect three diffuse light sources with respective total magnitudes of
  H=24.8, 25.5, and 25.9, plus a more compact object with a magnitude
  H=25.3. We discuss the significance of our detections and
  show that they are robust.}
{The three sources of diffuse light indicate an elongation along a
  north-east south-west axis, similar to that of the distribution of
  the central galaxies and to the X-ray elongation. This strongly
  suggests a history of merging events along this direction.
  While Guennou et al. (2012) found a roughly constant amount of
  diffuse light for clusters between z$\sim$0 and 0.8, we put in
  evidence at least a 1.5 magnitude increase between z$\sim$0.8 and
  2. If we assume that the amount of diffuse light is directly linked
  to the infall activity on the cluster, this implies that
  CL~J1449+0856 is still undergoing strong merging events.  }


\maketitle

\section{Introduction}

Clusters of galaxies are the final product of a long term hierarchical
accretion process, starting with the first matter density
concentrations visible in the Cosmic Microwave Background and
continuing until present days with still growing massive clusters such
as Coma.  We can define at least three benchmarks in the cluster
building process.

First, the actually favored cosmological $\Lambda$CDM model 
predicts that the first massive structures should appear around
z$\sim$2. More quantitatively, Romer et al. (2001) for example have
shown that structures hotter than 2~keV should not be found in a
$\Lambda$CDM universe with $\Omega_m$=0.3 approximately beyond this
redshift. Such temperatures of 2~keV can be considered as the lowest
possible values for clusters, group temperatures being closer to 1~keV
(e.g. Fig.~7 of Ulmer et al. 2005). The fact that the highest redshift
confirmed massive structures are located around z$\sim$2.1 (Gobat et
al. 2011) is in good agreement with this model (as long as other
massive structures are not detected at significantly higher
redshifts).
  
After having reached this initial (somewhat arbitrary) temperature of
2~keV, clusters continue to grow accreting matter from their
surrounding filaments. This growth is probably mainly driven by
accretion of already formed groups (e.g. Adami et al. 2005a). At
redshifts close to 2, such events could be considered as major mergers
between more or less equally massive structures, the impactor (the
infalling group) not being significantly less massive than the
impacted structure (the ``cluster''). At lower redshifts, the impacted
structure has gained mass, while infalling groups are still of low
mass, and we therefore reach a situation where fusions are becoming
minor.  This allows us to define a second crucial step in the cluster
history, which corresponds to the epoch (around z$\sim$1, see
Appendix A) where infalling groups are becoming significantly less
massive than the accreting cluster.

Clusters then mainly grow through minor fusions until they
(hypothetically) reach a virialized state. However, such virialized
clusters are very rare, even if a few are known at z$\sim$0
(e.g. A496, Durret et al.  2000). This shows that clusters have
generally not yet reached this third step.

We have therefore defined two main epochs in the cluster history:
their initial accretion phase between z$\sim$2 and 1, and their mature
growth at redshifts lower than 1.

There is accumulating evidence that cluster properties do not change very
significantly  during their mature evolution phase. Even if
their galaxy populations show variations during this period, with more
lenticular and fewer late type galaxies at z$\sim$0 than at z$\sim$1,
clusters do not seem for example to show strong dynamical evolution at
least up to z$\sim$0.5 (e.g. Adami et al. 2000), their galaxy red
sequence is already in place at least up to z$\sim$1.2 (e.g. Bremer et
al. 2006), and their diffuse light content remains stable up to
z$\sim$0.8 (Guennou et al. 2012).

The question is now to know if cluster properties evolve more
significantly at redshifts notably higher than 1. This is a new field
of investigation since very few clusters are known in this redshift
interval.  We propose in this paper to pave the way regarding the
cluster diffuse light content by trying to detect this cluster
baryonic component at z$\sim$2.1 in the structure discovered by Gobat
et al. (2011). 

Diffuse light becomes an important aspect in clusters
of galaxies as shown by recent studies (Gregg \& West 1998; Mihos et al. 
2005; Zibetti et al. 2005; Gonz\'alez et al. 2007; Krick \& Bernstein 2007; 
Rudick et al. 2010; Burke et al. 2012). It finds its origin at least partly 
in matter ejections
from galaxies during galaxy-galaxy or galaxy-cluster potential
interactions (see e.g. Adami et al.  2005b, or Dolag et al. 2010 for
recent simulations). More intense interactions should therefore lead
to a larger amount of diffuse light in clusters, and by the detection
of diffuse light we can evaluate the infalling activity on a
cluster. 

Galaxy evolution and ``pre-processing'' in groups, prior to their
  infall in clusters, also play an important role in the content of
  diffuse light in mature clusters (Fujita 2004; Rudick et
  al. 2006). The amount of pre-processing will depend on the
  characteristics of the infalling group, such as density and
  dynamical evolution stage (Da Rocha \& Mendes de Oliveira 2005, Da
  Rocha et al. 2008).

Guennou et al. (2012) used deep HST ACS data to perform detection of diffuse
light for 10 clusters between z=0.4 and 0.8 and did not detect a
significant variation of the cluster diffuse light content with
redshift. By applying a somewhat less sophisticated method, Burke et
al. (2012) extended this search to a sample of 6 clusters at z$\sim 1$
based on VLT/HAWK-I data, and found that IntraCluster Light (hereafter
ICL) only constituted $(1-4)$\% of the total cluster light.

Playing the same game at higher redshift is a difficult task, since
extremely deep images are required to beat the cosmological
dimming factor acting on extended sources following a (1+z)$^4$ law.  The
need for images with very low sky background is even more crucial at
redshifts significantly larger than~1, given the surface brightness of
the searched sources. This clearly speaks in favor of space-based
images. Finally, such a search  requires redder bands as we go to
higher redshifts, because the diffuse light usually detected in
clusters is rather red, and therefore probably not forming stars
(e.g. Adami et al. 2005b). Hence, we need to use photometric bands
sensitive to old stellar populations at the considered redshift
(typically redder than the rest-frame V band). At z$\ge$1, this
eliminates all visible bands, only leaving us with NIR bands.

Combining these three requirements drastically reduces the number of
clusters with suitable data available. The Gobat et al. (2011) z=2.07
structure is one of them. Being one of the very rare z$\ge$1.5
clusters with a measured X-ray temperature hotter than 2~keV, it has
been observed with the NICMOS HST near infrared camera in the F160W
band (close to the H band) with a $\sim$20 ksec exposure time. Such
data should make possible the detection of diffuse light sources
$only$ $\sim$five times brighter in flux than the typical Coma
sources, by applying the same wavelet technique as in Guennou et
al. (2012).

Section~2 describes the data. Section~3 describes the method
and presents our detections. Section~4 deals with the detection
robustness. Section~5 deals with the source redshift, and Section~6 is 
the discussion. We adopted the concordance 
cosmological model ($\Omega_{Lambda}$=0.74, $\Omega_{m}$=0.26). All
magnitudes are given in the AB system.

\section{Data}

As previously stated, we downloaded the HST NICMOS H band data for the
CL~J1449+0856 cluster (observing program 11174, PI: E. Daddi). The
data consist of two separate exposures of 7680 and 10240~sec in the
F160W filter. This cluster, with a center at $\alpha=222.3092^o,
\delta=+8.9404^o$ (see Gobat et al. 2011) also shows extended X-ray
emission, and is at a redshift of 2.07. Gobat et al. (2011) estimated
its mass to $(5.3-8)\ 10^{13}$ M$_\odot$.

The F160W filter at z=2.07 is more or less equivalent to the F814W
filter for clusters at z$\sim$0.56. This is close to the mean redshift
of the cluster sample analysed by Guennou et al. (2012) based on data
taken with the F814W filter and analysed with the same method we will
apply here. We are therefore in a very favourable situation to compare
the results of the present study with those of Guennou et al. (2012).

We used the NICMOS-pipelined mosaic images. This reduction is suitable
for our purposes because it only removes a constant sky level, and
therefore does not remove potential large scale diffuse light
sources. We then extracted a 134$\times$162 pixel$^2$ area centered on
the cluster position. This eliminated portions of the initial images
strongly affected by flat field residuals (see also Section 4).

We finally combined the two images to create a $\sim$18 ksec F160W
image. This was made with the Scamp and Swarp softwares (Bertin et
al. 2002; Bertin 2006) without background removal in order not to
erase diffuse light sources because of the Spline background estimate
method used by Swarp.

We note that the cluster does not occupy the same place on the NICMOS
detector in the two individual images, minimizing the effect of
  flat field and detector artifacts in our analysis.

\section{Method and results}

\subsection{The use of the OV\_WAV package}

\begin{figure}[!h]
  \begin{center}
    \includegraphics[width=3.40in,angle=0]{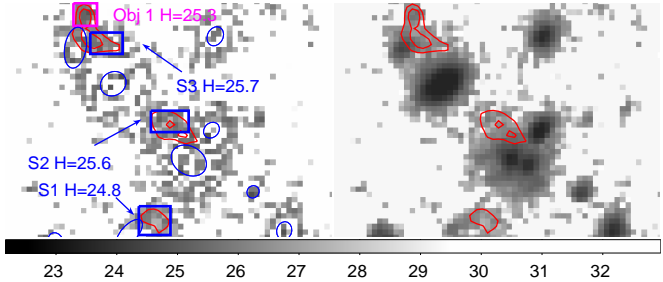}
    \caption{Right: F160W HST NICMOS image (100 kpc wide) of CL~J1449+0856. 
Left: residual image. Blue ellipses are SExtractor detected objects at a 
threshold of 1.4. Red contours are computed from the residual image. Contours 
start at the 2.5$\sigma$ level and have a 0.5$\sigma$ increment. We also give the 
H-band surface brightness scale and the integrated magnitudes of the detected sources.}
  \label{fig1}
  \end{center}
  \end{figure}

We applied the same wavelet-based method (the OV\_WAV package, see
e.g. Pereira 2003, and Da Rocha \& Mendes de Oliveira 2005)  to
detect diffuse light as applied to 10 clusters between z=0.4 and 0.8
by Guennou et al. (2012). A similar method was also used for the
  Coma cluster (z=0.023, Adami et al. 2005b) and Abell~2667 (z=0.233,
  Covone et al.  2006). We are therefore in the position of discussing
  the full cluster history, from z$\sim$0 to z$\sim$2.1. We briefly
  summarize below the method that we applied.  We only give the
  salient points of the technique in this section and refer the
  reader to Pereira (2003) and Da Rocha \& Mendes de Oliveira (2005)
  for more details. 

  OV\_WAV is a multi-scale vision model (e.g. Ru\'e \& Bijaoui 1997).
  After applying a wavelet transform to an observation of the targeted
  cluster, we identified the statistically significant pixels in the
  wavelet transform space (at the 5$\sigma$ level).  To define the
  objects, the selected pixels are then grouped in connected fields
  for each given scale. An inter-scale connectivity tree is then
  established and we identify each connected tree containing connected
  fields of significant pixels across three or more scales. We can
  then associate them with the objects.  An individual image is
  finally recovered for each identified object, with an iterative
  reconstruction algorithm.

In a first step, we detect small scale objects
in the sky image to produce our object image. 
We consider characteristic scales between 1 and 32 pixels in
wavelet space, adapted to the size of the NICMOS images. The object
image is then subtracted from the sky image to produce the residual
image (Fig.~\ref{fig1} left). This residual image includes both
hidden features which are typically too faint to satisfy the wavelet
first-pass thresholding conditions (fixed at 5$\sigma$) and the
diffuse light sources. We finally detect objects in this residual
image, on characteristic scales between 1 and 16 pixels, to create
a second object image.

In a second step we search for what we call the real ICL sources,
i.e. extended low surface-brightness features in the residual
image. These features are detected in this image by considering the
pixels where the signal is larger than 2.5$\sigma$ compared to an
empty area of the residual image (Fig.~\ref{fig1} shows the 2.5 and
3$\sigma$ levels as red contours). Each of these sources was visually
inspected before the removal of obvious numerical residuals of bright
saturated Galactic stars or defects due to image cosmetics. This
was done by subtracting a Sinc function (see Appendix B) with the orientation 
of the SExtractor detected objects. The amplitude of the Sinc function was
scaled to the SExtractor object magnitudes. We note that the three
diffuse light sources that we will discuss in the following were
detected as objects by OV\_WAV in the second object image, melted with
numerical residuals. This is why we used the residual image to locate
these diffuse light sources.

\subsection{Detected sources}

We display our results in Fig.~\ref{fig1}. Objects that are
  wavelet-detected in the first pass are compact and also nearly all
  SExtractor-detected.  The residual image (Fig.~\ref{fig1} left)
  includes three diffuse light sources (S1, S2, S3) and probably a
  compact low signal to noise object (indicated as Obj~1).  We also
  show the blue rectangles in which we measured the magnitudes for S1,
  S2, S3, and Obj~1.  S1, S2, and S3 have respective total magnitudes of
  H=24.8$\pm$0.13, 25.5$\pm$0.21, and 25.9$\pm$0.34 (see next
  section). Their mean surface brightness are 25.95, 26.0, and 26.3. 
  This amount of diffuse light in CL~J1449+0856 roughly
    corresponds to a L$^*$ field galaxy at z$\sim$2 (e.g. Ilbert et
    al. 2005). This is a similar amount to what we found in the Coma
    cluster (Adami et al. 2005b).  Obj~1 is a compact object with a
  total magnitude of H=25.3.

We note that the S1 source is located 2.4 arcsec from the optical
center of CL~J1449+0856.  At z=2.07, this translates to almost 20 kpc,
typical of the clustercentric distances found for the diffuse light
sources of Guennou et al. (2012). This also means that such a study is
impossible with classical ground-based data where the typical seeing
is of the order of 1~arcsec.

\section{Detection robustness}

\subsection{Simulations}

\begin{figure}[!h]
  \begin{center}
    \includegraphics[width=3.40in,angle=0]{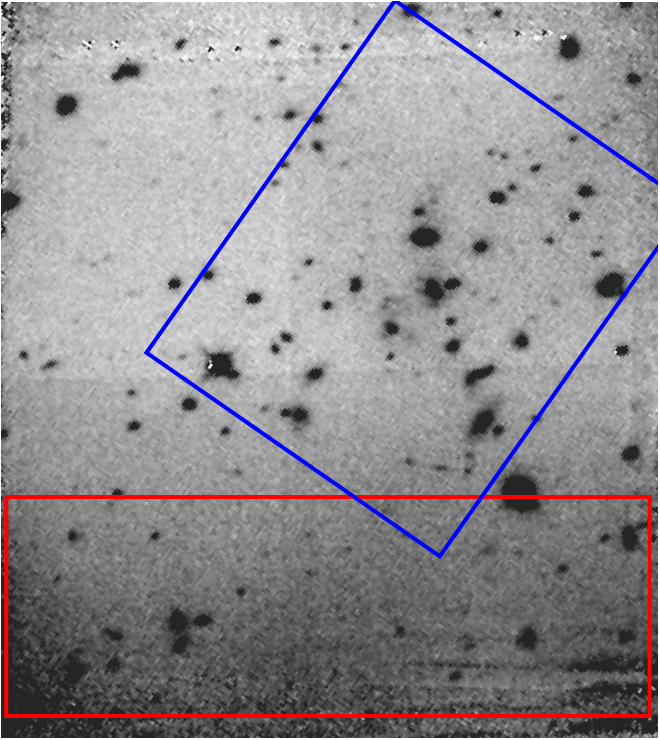}
    \caption{H-band NICMOS image of CL~J1449+0856. The blue rectangle is the 
region centered on the cluster itself, and the red rectangle is the 
region with the strongest flat field residuals.}
  \label{regions}
  \end{center}
  \end{figure}

\begin{figure}[!h]
  \begin{center}
    \includegraphics[width=2.20in,angle=270]{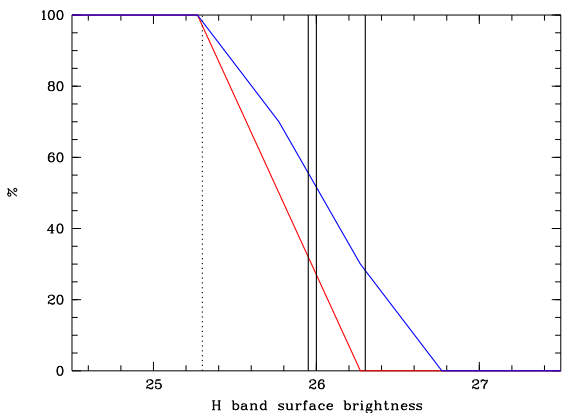}
    \caption{Percentage of recovered diffuse light sources as defined in the text
versus H-band surface brightness. Blue line: percentage in the cluster region. 
Red line: percentage in the high noise region (see Fig.~\ref{regions}). The three 
vertical solid lines
show the surface brightness of the three real diffuse light sources. The 
vertical dotted line is the surface brightness of Obj1 as shown in Fig.~\ref{fig1}.}
  \label{eff}
  \end{center}
  \end{figure}

\begin{figure}[!h]
  \begin{center}
    \includegraphics[width=3.20in,angle=0]{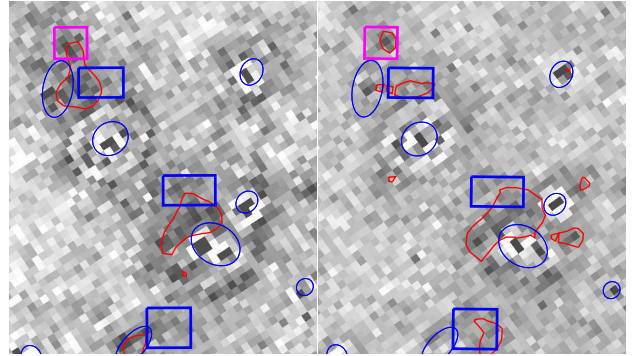}
    \caption{Residual images computed on the individual exposures
      (left: 7680~sec, right: 10240~sec). The red contours show the 2$\sigma$
detection level. The blue ellipses are the SExtractor-detected objects. The blue
rectangles show the locations of the  diffuse light sources detected in the summed
image (see Fig.~\ref{fig1}).}
  \label{fig3}
  \end{center}
  \end{figure}

  We first address the question of the theoretical detection limit of
  our method with the data in hand. Considering the simulations by
  Guennou et al. (2012), OV\_WAV should not allow us to detect
  Coma-like 2.5$\sigma$ diffuse light sources at z greater than 0.8 in
  10~ksec F814W HST ACS exposures. Assuming a (1+z)$^4$ dimming
  factor, the same restframe wavelength, and an exposure of 18~ksec,
  this implies that we could only detect sources 1.7
  magnitude/arcsec$^2$ brighter than Coma-like sources at z=2.07 with
  F160W at the 2.5$\sigma$ level, and sources 1.2 magnitude/arcsec$^2$
  brighter than Coma-like sources at z=2.07 with F160W at the
  2$\sigma$ level. Note however that these estimates are based on
    two different instruments (ACS versus NICMOS) and related to
    different redshifts and observed wavelength ranges (already formed
    clusters at z$\leq$1 versus a young cluster just at the beginning
    of its formation at z$\sim$2).

    We therefore redid simulations similar to those of Guennou et
    al. (2012). We simulated sources of uniform surface brightnesses
    with 10 pixel diameters, which we scaled to different surface
    brightnesses before inserting them into the NICMOS images of
    CL~J1449+0856.  This size is typical of the sources we detect in
    CL~J1449+0856.  As in Guennou et al. (2012), we performed this
    exercice 10 times for each simulated surface brightness. This
    allowed to quantify our ability to detect sources of diffuse
    light, as well as the purity of our detections (how many fake
    sources we are detecting). 

Knowing that relatively strong flat field residuals are
affecting the external parts of the HST NICMOS images, we can ask the
question of the contribution of this effect to the diffuse light
detection. We therefore performed the described simulations in
  two areas. The first one is centered on the cluster and the second
  one is located in the image region which is the most affected by
  flat field residuals (see Fig.~\ref{regions}).

  We show the results in Fig.~\ref{eff}. We see that we are able to
  detect 100$\%$ of the diffuse light sources down to a brightness of
  H$\sim$25.2~mag~arcsec$^{-2}$. The detection percentage then
  decreases rapidly when considering the image region that is
  strongly affected by flat field residuals to reach a null efficiency
  for brightness of H$\sim$26.2~mag~arcsec$^{-2}$. On the other hand,
  we see that we can detect sources down to
  H$\sim$26.8~mag~arcsec$^{-2}$ in the cluster region. These results
  are in good agreement with our detections. Obj1 (see
  Fig.~\ref{fig1}) is located just at the limit of the full detection
  range, while the three diffuse light sources are located in the
  decreasing part of the blue curve in Fig.~\ref{eff}. We note that
  the faintest diffuse light source would not have been detected if it
  had been located in the high flat field residual region.
  Fig.~\ref{eff} also implies that we possibly miss diffuse light
  sources with brightnesses fainter than H$\sim$26.8~mag~arcsec$^{-2}$
  in our detections, so our estimate of the total amount of diffuse
  light in CL~J1449+0856 is only a lower limit.

  Finally Fig.~\ref{eff} shows that our results are probably only
  weakly affected by fake detections because no such detections occur
  for very faint brightnesses (where we theoretically are unable to
  detect our simulated sources). This means that our purity is higher
  than 90$\%$: less than 1 fake detection is occuring among our 10
  simulations.

  With these simulations, we were also able to estimate  the typical
  uncertainty on the magnitudes of S1, S2, and S3, computed as the
  1$\sigma$ variation of the recovered surface brightnesses. This allowed us 
  to estimate the magnitude uncertainties given in Section 3.2.

\subsection{Individual exposures}

Another test is to try to detect diffuse light sources in each of the
two individual NICMOS images in order to estimate the robustness of
our detections.  Since the cluster does not fall at the same place on
the detector in the two exposures, flat field problems should not
affect the detection of diffuse light in the same way. We do not
expect to detect S1, S2, and S3 at better than the 2$\sigma$ level in
individual exposures but the fact that they could be visible in each
of the two individual images would give confidence in our summed-image
detections.  We show in Fig.~\ref{fig3} the residual images
  obtained for the two individual exposures. The 2$\sigma$ level
  detections are similar to the results obtained with the 18~ksec
  combination.  S1 is not significantly detected in the 7680~sec
  exposure, but appears in the 10240~sec exposure. S2 also seems  to
  appear in both exposures, while S3 and Obj1 are detected in the two
  individual images. Detection levels are low (2$\sigma$ versus
  2.5$\sigma$ for the combined image), as expected, but this gives
  confidence in our detections performed on the 18~ksec combined
  image.

\section{Redshift of the diffuse light sources}

\begin{figure}[!h]
  \begin{center}
    \includegraphics[width=2.40in,angle=270]{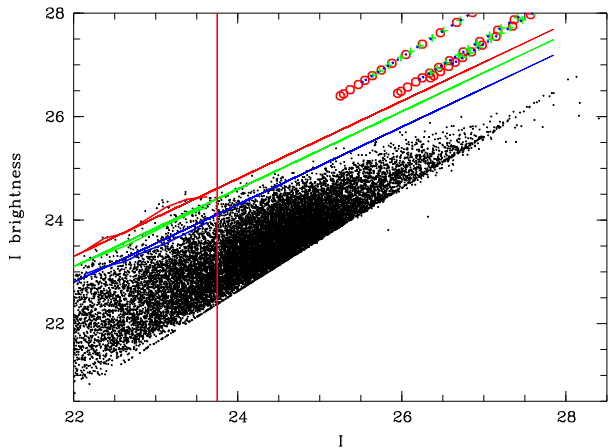}
    \caption{Mean F814W surface brightness (see text) versus I-band 
      (F814W) total
      magnitude. Black dots: Cosmos (Scoville et al. 2007) HST ACS
      detected objects with a 68$\%$ photometric redshift error lower
      than 0.2 and a photometric redshift between 0.1 and 6. Only one
      object over 100 is shown for clarity. Red circles, green
      crosses, and blue dots: the three detected diffuse light sources
      in CL~J1449+0856 assuming different redshifts and different
      types for the three sources. This allowed us to translate H-band
      to I-band magnitudes. The vertical red line is the assumed
      Cosmos completeness total magnitude in terms of surface brightness
      (I=23.75: see text). The inclined straight lines show the fits of
      the Cosmos detection limits (broken lines) computed for total magnitudes
      I$\leq$23.75 (red: 3$\sigma$, green: 2.5$\sigma$, blue:
      2$\sigma$): see also text.}
  \label{cosmos}
  \end{center}
  \end{figure}

  The next crucial step is to estimate the redshift of the detected
    diffuse light sources. It is obviously impossible with the
    instrumentation presently available to measure a spectroscopic
    redshift for these sources. It is also impossible to compute a
    photometric redshift with the single available photometric
    band. We will therefore rely on statistical arguments to show that
    the three detected diffuse light sources are probably members of
    the CL~J1449+0856 cluster.  Several objects could mimic cluster
    diffuse light sources, in particular: (1) relatively bright and single
    extended objects on the line of sight, (2) associations of very faint
    objects.

\subsection{Single galaxies along the line of sight?}

    In order to test hyphothesis (1), we considered the Cosmos
    HST-ACS catalog (Scoville et al. 2007). We computed 
    the mean surface brightness using the Cosmos aperture magnitude measured
    inside a 0.3 arcsec radius area.  

    - This catalogue extends
    over $\sim$2 deg$^2$, provides objects down to I$\sim$26.5 (total
    magnitude), and
    theoretically includes all objects (including the faintest possible
    surface brightness objects) down to I$\sim$23.75 (see
    Fig.~\ref{cosmos}). The star-galaxy separation is efficient down
    to I$\sim$24. Objects fainter than this limit become so small that
    even the Cosmos high resolution images start to be unable to resolve them.
    Besides the consequences on our surface brightness computation, this
    also underlines our inability to select pure galaxy samples at these
    magnitudes.

    - The Cosmos survey does not include H-band measurements, so we
    had to compute F814W~-~F160W colors for different redshifts and
    different spectral types in order to be able to put our three
    diffuse light sources on the same plot. This was done with the
    LePhare tool (Ilbert et al. 2006) between redshifts of 0.1 and 6
    and for three different spectral types (Elliptical1: red circles,
    Sa0: blue dots, and Sd0: green crosses in Fig.~\ref{cosmos}) among
    the Cosmos templates.

  - We assumed the Cosmos galaxy catalog complete in terms of
    surface brightness and not polluted by stars down to I$\sim$23.75
    (total magnitude). In other words, we assumed that all objects
  brighter than I$\sim$23.75 are detected in the Cosmos survey
  whatever their surface brightness, and that we can separate
    stars from galaxies. We then computed the 3$\sigma$,
  2.5$\sigma$, and 2$\sigma$ Cosmos detection limits in terms of
  surface brightness down to I$\sim$23.75. In other words, for a given
  total I-band magnitude, we computed the maximal mean surface
  brightness allowing the detection of 99.7$\%$ of the galaxies
  (i.e. the 3$\sigma$ level: red broken line in Fig.~\ref{cosmos}), of
  98.75$\%$ of the galaxies (i.e. the 2.5$\sigma$ level: green
  broken line in Fig.~\ref{cosmos}), and of 95.45$\%$ of the 
    galaxies (i.e. the 2$\sigma$ level: blue broken line in
  Fig.~\ref{cosmos}). Finally, we fit straight lines on these broken
  lines to roughly estimate the surface brightness limitations
  at fainter total I-band magnitudes to detect galaxies at the
  3$\sigma$, 2.5$\sigma$, and 2$\sigma$ levels, where objects are
    too small to be resolved in the considered images. Note that these
    lines are in fact upper limits, since one would expect that as
    magnitudes become fainter, the surface brightness limit cannot
    continue to increase linearly.

    Fig.~\ref{cosmos} shows that less than 0.3$\%$ of the Cosmos
    galaxies could be faint enough in terms of surface brightness to
    mimic our three diffuse light sources. Considering the density of
    known Cosmos galaxies on the sky and the spatial extension of our
    diffuse light sources, we conclude that we only have 15$\%$ of
    chances to have a Cosmos-like galaxy on the line of sight
    explaining S1, 8$\%$ of chances for S2, and 7$\%$ of chances for
    S3. The probability for all three diffuse light sources to be
    explained by Cosmos-like galaxies on the line of sight is therefore
    lower than 0.1$\%$. 

    The last possibility in the hypothesis (1) framework would be to
    have detected large and extremely low surface brightness galaxies
    on the line of sight, faint enough in surface brightness not to be
    detected by the Cosmos survey. To test this, we used the field low
    surface brightness galaxy catalog detected in Adami et
    al. (2006). In a 30$\times$30~arcmin$^2$ empty sky region, we
    detected 29 galaxies with B$\leq$25.5 and B surface brightness
    $\leq$27.2~mag~arcsec$^{-2}$. Assuming a B-I color varying between
    1 and 6 (from LePhare computations) between z=0 and 4, the
    probability to have such a low surface brightness galaxy in a 10
    $\times$ 10 arcsec$^2$ area (the presently considered region of
    interest including S1, S2, and S3) varies between 0.2\% and
    8$\%$. The probability to have 3 such galaxies is therefore lower
    than 0.05$\%$.

\subsection{Associations of very faint objects along the line of sight?}

    Associations of very faint objects along the line of sight (hypothesis 2)
    could  also explain our diffuse light sources. They could
    be  compact classical objects that are spatially correlated. In order to
    test this, we considered the Hubble Ultra Deep Field (Beckwith et
    al. 2006), providing an object catalog down to i$\leq$29.75 (on a
    more reduced area than the Cosmos region). Given the spatial
    density of HUDF objects, only 0.3, 0.17, and 0.15 objects could
    statistically be located in the S1, S2, and S3 areas.  This is
    clearly not enough to explain an association of objects mimicking our
    diffuse light sources. The only way to reach such a result would
    be to have exceptionally detected 3 extremely distant groups of
    galaxies in a $\sim 10\times 10$~arcsec$^2$ area. Even with the most favourable
    cluster-cluster angular correlation functions (e.g. Durret et
    al. 2011), the probability to find 3 groups hotter than 2 keV
    (Romer et al.  2001) in a 10 $\times$ 10 arcsec$^2$ area to
    explain S1, S2, and S3 is completely negligible
    ($\leq$10$^{-14}$).

    We can therefore conclude quite securely that the best possibility
    to have detected three such low surface brightness objects in 
    such a small area is to have detected diffuse light sources related to the
    CL~J1449+0856 cluster.

\section{Discussion}

\begin{figure}[!h]
  \begin{center}
    \includegraphics[width=2.5in,angle=270]{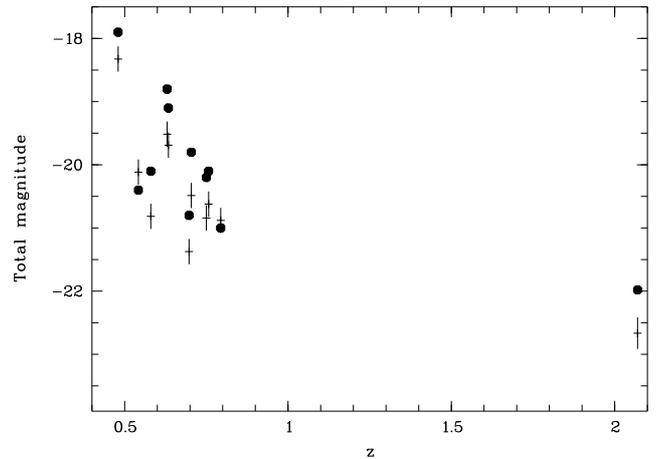}
    \caption{Total absolute magnitude of the detected diffuse light
      (F814W for the 10 Guennou et al.  (2012) clusters to the left, and F160W
      for CL~J1449+0856 to the right). The F160W filter at z=2.07 is 
       equivalent to the F814W filter for clusters at z$\sim$0.56. Filled 
circles: raw values,
      crosses: mass normalized values with Fig.~10 of Guennou et
      al. (2012). Error bars are only plotted for the mass normalized 
values. The Guennou et al. (2012) points were assumed to be plagued by a 
$\pm$0.2 magnitude error, as suggested by Fig.10 of this paper.}
  \label{fig4}
  \end{center}
  \end{figure}

As already stated in this work, the general process of the
  cluster building history starts by the growth of an initial mass
  fluctuation. This seed is then fed by infalling matter along the
  cosmological filaments, mainly in the form of already formed groups
  of galaxies (in which galaxies are pre-processed). This induces
  interactions between galaxies, and between infalling galaxies and the
  cluster potential. The intensity and frequency of these interactions
  depend on many parameters, such as the stage of dynamical evolution
  of the infalling groups and of the growing cluster.  Interactions
  induce matter ejections from galaxies through for example dynamical
  friction or tidal disruptions. The signatures of such processes are
  then directly detectable through intracluster diffuse light source
  detections. Their distribution in a given cluster (assuming that the
  survival time of these sources is not too long) is therefore
  expected to give direct indications on the zones where galaxy
  interactions have taken place, and on the directions from which
  infall has occured on to the cluster.

We applied this technique to the CL~J1449+0856 cluster, comparing the 
indications given by intracluster diffuse light sources to other indicators:
the position of the central galaxies of CL~J1449+0856 clearly indicates
a north-east south-west axis. The same orientation is visible in the
S1, S2, S3 distribution.  This elongation is similar to that found in
X-rays from XMM-Newton data by Gobat et al. (2011, see the white
ellipse on their Fig.~1, left) and therefore suggests a history of 
merging events along this direction.

Similarly, the amount of diffuse light detected in a cluster will also
give insight on the intensity of the infall process undergone by the
cluster.  We show in Fig.~\ref{fig4} the total absolute magnitude (in
F814W rest frame for the Guennou et al. sample) of the diffuse light
detected in CL~J1449+0856 and in the Guennou et al. (2012) clusters as
a function of redshift. We give the raw values as well as the mass
normalized values, as in Guennou et al.  (2012, Fig. 10). We note
  that these mass normalized values are computed using the relation
  between mass and amount of diffuse light given in Fig.10 of Guennou
  et al. (2012) to take into account the natural tendency of a massive
  cluster to exhibit larger amount of diffuse light compared to a
  lower mass cluster.  While Guennou et al.  (2012) found a more or
less constant amount of diffuse light for clusters between z$\sim$0
and 0.8, we put in evidence a brightening by at least 1.5 magnitude
between z$\sim$0.8 and 2. If we assume that the amount of diffuse
light is directly linked to the infall activity on a given cluster, as
discussed in Guennou et al. (2012), we have here a clear sign of the
intense merging processes ongoing in CL~J1449+0856, as expected for
such a young and distant cluster.

\begin{acknowledgements}
The authors thank the referee for useful and constructive comments and
Johan Richard for his help. We acknowledge long term financial support from CNES.

\end{acknowledgements}

\appendix

\section{Infalling activity on clusters in the Millennium simulation}

\begin{figure*}[!h]
  \begin{center}
    \includegraphics[width=5.0in,angle=270]{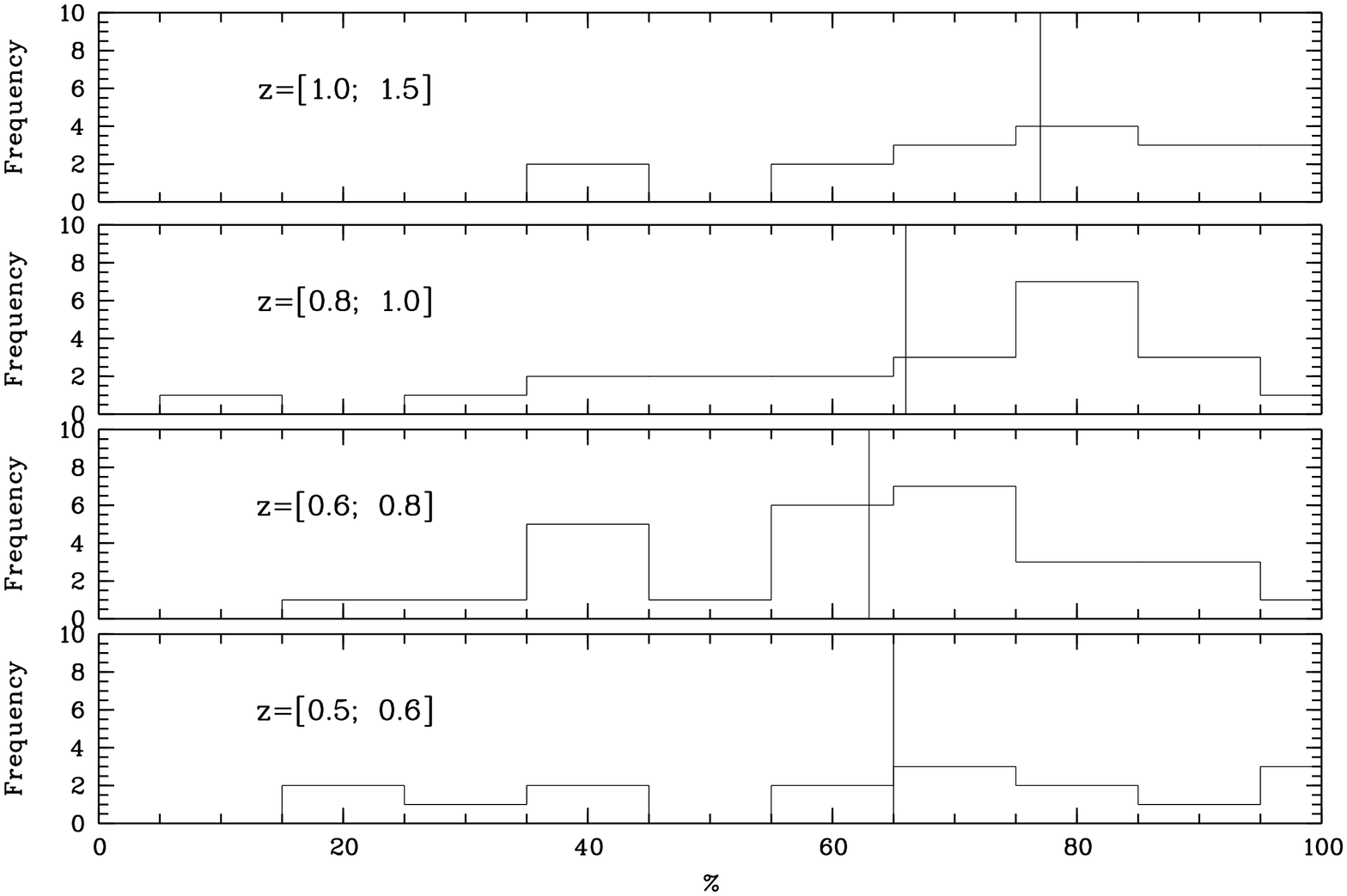}
    \caption{Histograms of the mass ratios of the most massive
      infalling structure to the main structure in the Millennium
      simulation given as percentages, in four redshift bins. The
      vertical lines show the mean values for each redshift bin. A
      percentage of 100 means that the most massive infalling
      structure has the same mass as the main structure.}
  \label{fig0}
  \end{center}
  \end{figure*}

The approximate redshift where infalling groups are becoming less
massive than the accreting cluster can be evaluated using the
Millennium simulation (Springel et al. 2005), based on a $\Lambda$CDM
universe with $\Omega_m$=0.3 and $\Omega_{\Lambda}$=0.7 (close to the
concordance model of Dunkley et al. 2009).  For haloes more massive
than 4~10$^{13}$ M$_{\odot}$, we computed in four redshift intervals
the ratio (expressed in percentages) between the halo mass and the maximum 
mass of the other haloes in a 2.5~Mpc radius (a typical virial radius for 
clusters) and within 0.01 in redshift. Such neighbors are assumed here to be
structures that will be accreted in the future. We show in
Fig.~\ref{fig0} that a cut seems to appear at z$\geq$1 where infalling haloes
are closer to the value of the main system mass. The position of this cut is 
also in good agreement with the expectations of Ulmer et al. (2009), based
on the typical relaxation time of a galaxy falling onto a typical
cluster. This will obviously have to be confirmed on larger numerical 
simulations in the future given the relatively small number of massive 
structures present in the Millenium simulations.

\section{Sinc subtraction in the residual image}

\begin{figure}[!h]
  \begin{center}
    \includegraphics[width=2.0in,angle=270]{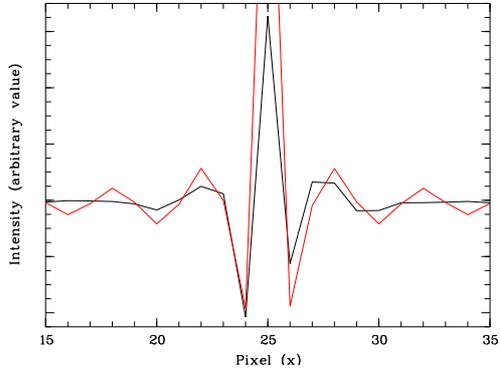}
    \caption{Black curve: numerical artifacts along the image x-axis coming 
from the brightest star in the CL~J1449+0856 field of view 
($\alpha$=14:49:14.820; $\delta$=+08:56:12.70) after the removal of the object 
image from the sky image (see Section 3). Red curve: Sinc modelisation.}
  \label{Sinc}
  \end{center}
  \end{figure}

In order to better remove the numerical artifacts in the residual
image, we subtracted from this image a SinC function centered on each
object and scaled to the object magnitude. In order to test the
validity of this approach, we show in Fig.~\ref{Sinc} (black curve)
the numerical artifacts coming from the brightest star in the
CL~J1449+0856 field of view ($\alpha$=14:49:14.820;
$\delta$=+08:56:12.70) after the removal of the object image from the
sky image (see Section 3). These artifacts can be modeled by a SinC function
(red curve). The subtraction of this analytical curve will therefore
remove the negative and positive annuli circling the central peak, at
the cost of having an oversubtraction of this residual central peak.

\end{document}